# Nature of point defects in bulk hexagonal diamond

Ling Zhu,[1,2] Xuanxuan Zhang,[1,2] Guliqinayi Alimu,[1,2] Chen-Min Dai,[1,2,*] Chunlan Ma,[1,2] and Zenghua Cai[1,2,*]

[1]*Key Laboratory of Intelligent Optoelectronic Devices and Chips of Jiangsu Higher Education Institutions, School of Physical Science and Technology, Suzhou University of Science and Technology, Suzhou, 215009, China*

[2]*Advanced Technology Research Institute of Taihu Photon Center, School of Physical Science and Technology, Suzhou University of Science and Technology, Suzhou, 215009, China*

[*]Contact author: daichenmin@usts.edu.cn; zhcai@usts.edu.cn

**Abstract**

Hexagonal diamond (HD), an exotic carbon allotrope recently synthesized in bulk form, exhibits superior mechanical properties compared to cubic diamond (CD) and holds promise for advanced industrial and quantum applications. Using first-principles calculations, we systematically investigate intrinsic defects, extrinsic dopants, and defect complexes in HD. Our study shows that $V_C$ dominates intrinsic conductivity, while $C_i$ is unstable. Among extrinsic dopants, boron acts as a benign acceptor enhancing p-type conductivity, whereas nitrogen and phosphorus serve as effective donors for n-type conductivity. Group II and Group IV dopants, however, introduce high formation energies or neutral charge states with limited impact. Furthermore, $V_C$, $Mg_C$ and XV defect complexes display multiple spin and charge states within the HD band gap, highlighting their potential as color centers for hosting qubits. These results not only clarify the defect physics of HD but also demonstrate its broader implications for conductivity engineering and quantum technologies.

## I. INTRODUCTION

Cubic diamond (CD) is widely recognized as the hardest known material, distinguished by its exceptional mechanical properties, including high compressive strength, bulk modulus, and Young's modulus[1-3]. Often characterized as the ultimate semiconductor, CD possesses unique chemical and optical characteristics that make it indispensable for a broad spectrum of industrial and scientific applications[1, 4]. It is extensively utilized as a superabrasive for precision cutting and grinding, as a high-performance heat sink for electronic components, and in various medical fields for prostheses and biological sensing[1, 2, 5]. Furthermore, its capacity to host optically active point defects with long spin coherence times has established it as a premier platform for quantum technologies, including quantum computing registers and highly sensitive magnetic field sensors[6, 7].

In addition to the cubic structure, diamond also has a hexagonal configuration, i.e., hexagonal diamond (HD), also known as lonsdaleite. This exotic carbon allotrope, historically associated with meteorite impacts, is characterized by an ABAB stacking sequence[1, 3, 8]. Although traditionally difficult to isolate as a discrete phase, recent breakthroughs have enabled the synthesis of millimeter-sized, phase-pure bulk HD by compressing highly oriented pyrolytic graphite (HOPG) or glassy carbon under high-pressure, high-temperature (HPHT) conditions[1, 2]. Both theoretical and experimental studies confirm that HD exhibits superior mechanical properties compared to CD, including higher indentation strength, greater stiffness, and a maximum compressive strength that can exceed CD by more than 30%[1, 3, 9]. Additionally, bulk HD demonstrates remarkable thermal stability and oxidation resistance, making it a compelling candidate for next-generation industrial applications in extreme environments[1].

Point defects play a critical role in semiconductors such as CD and HD, as they strongly affect the physical properties of semiconductors and have a decisive impact on their performance in applications[10]. Researches into the point defects in CD have been extensive[7, 11-13], with particular emphasis on color centers like the nitrogen-vacancy (NV) center, which functions as a robust solid-state qubit[6, 7, 14]. In contrast, investigations of point defects in HD remain limited, with studies largely restricted to preliminary examinations of the NV center and the $N_2V$ center[15, 16]. While CD currently serves as the standard for quantum emitters, the unique crystal field and lower symmetry of the hexagonal lattice are predicted to offer new defect configurations and tunable electronic structures that are not accessible in the cubic phase[15, 16]. Consequently, there is an urgent need for systematic research on HD point defects to fully characterize their potential for advanced sensing and quantum information processing[1, 2].

This study employs first-principles calculations to systematically investigate point defects in HD. We begin by examining intrinsic defects, including simple point defects (vacancies and interstitials) as well as defect complexes, specifically the divacancy (VV) center. Then, a comprehensive analysis of extrinsic doping across four key chemical groups is presented, encompassing point defects (substitutions and interstitials) and defect complexes, particularly the XV centers. Here, X denotes extrinsic impurity elements: (i) alkaline earth metals (Mg, Ca, Sr); (ii) Group III elements (B, Al, Ga); (iii) Group IV elements (Si, Ge, Sn, Pb); and (iv) Group V elements (N, P, As).

## II. THEORETICAL DETAILS

### A. Details of the calculations

All the first-principles calculations were performed using density functional theory (DFT) as implemented in the VASP code[17]. Defect property simulations employed a single Γ point (1×1×1 Monkhorst-Pack k-point mesh)[18] and a 432-atom HD supercell. A plane-wave cutoff energy of 420 eV was adopted, which is 50 eV higher than that used in previous work[13] and yields the converged results. Considering the substantial computational consumption, structural relaxations of the point defects were conducted based on the Perdew-Burke-Ernzerhof (PBE) exchange correlation functional within the generalized gradient approximation (GGA)[19] until the residual forces on all atoms are less than 0.01 eV/Å. The screened hybrid functional of the Heyd, Scuseria, and Ernzerhof (HSE06)[20] was used for all the total energy calculations of point defects with the fraction of exact exchange set to 0.32 ($\alpha = 0.32$). This setting yields a band gap of 4.62 eV for the 432-atom HD supercell, in excellent agreement with the 4.68 eV obtained from GW calculations[21]. Benchmark calculations of C vacancy defect in -1 charge state demonstrate that the total energy difference between structures relaxed by HSE06 and PBE is less than 0.01 eV as shown in Table SI of the Supplemental Material[22]. Spin polarization was included to determine the ground electronic states of defects, particularly the complex defects[23]. Defect properties were simulated using the supercell model, following the procedure described in DASP code[24]. The correction for the image charge interaction was implemented using the Freysoldt-Neugebauer-Van de Walle (FNV) scheme[25]. The static dielectric constant was set to 6.05, corresponding to the average of the values for light polarized parallel and perpendicular to the c-axis[26].

### B. Details of the supercell model calculations

According to the supercell model[24], the formation energy $\Delta H_f(\alpha, q)$ of a point defect $\alpha$ in the charge state $q$ (introduced by tuning the number of electrons in the defect system) can be calculated as follows:

$$\Delta H_f(\alpha, q) = E(\alpha, q) - E(\text{perfect}) - \sum_i n_i(E_i + \mu_i) + q[E_{VBM}(\text{perfect}) + E_F] + E_c, \quad (1)$$

where $E(\alpha, q)$ and $E(\text{perfect})$ denote the total energies of the HD supercell with and without defects, respectively. $n_i$ is the number of atoms *i* removed from ($n_i < 0$) or added to ($n_i > 0$) the supercell. $E_i$ is the energy per atom of the element *i* (e.g., the intrinsic elements C or extrinsic elements Mg, B, Si, N) in its elemental phase. $\mu_i$ is the chemical potential of the element *i*. $E_{VBM}(\text{perfect})$ is the valence band maximum (VBM) level of the perfect supercell. $E_F$ is the Fermi level referenced to VBM, and the Fermi level shifts up from the VBM to the conduction band minimum (CBM) when $E_F$ increases from zero to the band gap value. $E_c$ is the correction energy for the image charge interaction[25].

As defined in Eq. (1), the defect formation energy depends on the chemical potentials of the atoms removed or added during the defect formation. The chemical potentials are limited by a series of thermodynamic conditions. First, the leftover of the pure elemental phases of the intrinsic (C) or extrinsic elements (Mg, B, Si, N, etc.) must be avoided, so $\mu_i$ ($i$ represents these elements) should satisfy:

$$\mu_i < 0. \quad (2)$$

Meanwhile, HD should reach an equilibrium state, so $\mu_{\text{C}}$ should satisfy:

$$\mu_{\text{C}} = \Delta H_f(\text{HD}). \quad (3)$$

Since graphite is the most thermodynamically stable allotrope of C, $\Delta H_f(\text{HD})$ is positive. Therefore, $\mu_{\text{C}}$ is set to zero for all intrinsic point defect calculations. For extrinsic doping, the formation or coexistence of the competing secondary phases composed of the C and extrinsic elements should be avoided. For example, in the case of Mg doping, the binary competing phases $Mg_2C_3$, $MgC_2$, $Mg_3C$ and MgC should be avoided. Therefore, the chemical potentials of their component elements should satisfy:

$$2\mu_{\text{Mg}} + 3\mu_{\text{C}} < \Delta H_f(\text{Mg}_2\text{C}_3), \quad (4)$$

$$\mu_{\text{Mg}} + 2\mu_{\text{C}} < \Delta H_f(\text{MgC}_2), \quad (5)$$

$$3\mu_{\text{Mg}} + 2\mu_{\text{C}} < \Delta H_f(\text{Mg}_3\text{C}), \quad (6)$$

$$\mu_{\text{Mg}} + \mu_{\text{C}} < \Delta H_f(\text{MgC}). \quad (7)$$

The allowed chemical potential ranges of C and Mg that stabilize the pure-phase HD are limited by these equations and inequations. Key competing secondary phases and their formation energies are listed in Table SII of the Supplemental Material, while the allowed chemical potential ranges of C and extrinsic elements are provided in Table SIII[22]. From Table SII, we can find that the formation energies of most competing phases are positive, resulting in only minor differences between C-rich and C-poor conditions. Consequently, defect properties vary negligibly between these two conditions, and only the C-rich condition is considered in this work.

Additionally, transition energy level $\varepsilon(q/q')$ can be calculated as the $E_F$ at which the defect α in the charge state $q$ has the same formation energy as in the charge state $q'$. It can be expressed as:

$$\varepsilon(q/q') = \left(\Delta H_f(\alpha, q) - \Delta H_f(\alpha, q')\right)/(q' - q). \tag{8}$$

Table I. The calculated binding energies $E_b$ for all neutral defect complexes considered in this work, along with the corresponding neutral formation energies ($\Delta H_f$) of both defect complexes and isolated point defects.

| Type | $\Delta H_f$(eV) | Type | $\Delta H_f$(eV) | Type | $\Delta H_f$(eV) |
|---|---|---|---|---|---|
| $V_C$ | 7.10 | $Mg_C$ | 11.80 | $Ca_C$ | 17.03 |
| $Sr_C$ | 21.27 | $B_C$ | 1.68 | $Al_C$ | 8.71 |
| $Ga_C$ | 9.17 | $Si_C$ | 4.31 | $Ge_C$ | 6.89 |
| $Sn_C$ | 13.74 | $Pb_C$ | 17.44 | $N_C$ | 3.80 |
| $P_C$ | 6.65 | $As_C$ | 10.26 | | |
| Type | $\Delta H_f$(eV) | $E_b$ (eV) | Type | $\Delta H_f$(eV) | $E_b$ (eV) |
| $VV_a$ | 10.53 | 3.67 | $VV_b$ | 10.46 | 3.74 |
| $MgV_a$ | 9.92 | 8.98 | $CaV_a$ | 13.57 | 10.56 |
| $SrV_a$ | 17.12 | 11.25 | $BV_a$ | 6.24 | 2.54 |
| $AlV_a$ | 8.61 | 7.2 | $GaV_a$ | 9.04 | 7.23 |
| $SiV_a$ | 6.84 | 4.57 | $GeV_a$ | 8.51 | 5.48 |
| $SnV_a$ | 11.46 | 9.38 | $PbV_a$ | 14.24 | 10.3 |
| $NV_a$ | 6.28 | 4.62 | $PV_b$ | 5.30 | 8.45 |
| $AsV_b$ | 7.47 | 9.89 | | | |

### C. Details of the binding energy calculations

In order to calculate the binding energies ($E_b$) of defect complexes, the following method is used:

$$E_b = \sum \Delta H_f(\alpha_i, q) - \Delta H_f(\alpha_c, q), \tag{9}$$

where $\Delta H_f(\alpha_i, q)$ represents the formation energy of isolated point defect $\alpha_i$ and $\Delta H_f(\alpha_c, q)$ is the formation energy of the defect complex. The calculated binding energies for all the neutral defect complexes considered in this work are listed in Table I. The results show that all the binding energies for the defect complexes are positive, indicating that all the defect complexes are more stable than the corresponding isolated point defects.

## III. RESULTS AND DISCUSSION

### A．Intrinsic point defects and defect complexes

Bulk hexagonal diamond crystallizes in the space group *P*$6_3$/*mmc*[1] and contains only one inequivalent C atom in its lattice as shown in Fig. 1(a). The lattice site of this inequivalent C atom is commonly called hexagonal site (h site). Depending on orientation, the C-C bonds in HD can be classified into two types: (i) bonds along axis direction, i.e., [0001] direction; (ii) bonds along basal direction. As shown in Fig. 1(a), after structural relaxation, the bond length along the axis direction is 1.55 Å, while the bond length along the basal direction is 1.53 Å. From the symmetry perspective, intrinsic point defects in HD consist of two types: C vacancy ($V_C$) and C interstitial ($C_i$). For $V_C$, only one type exists since only one inequivalent C atom in HD. For $C_i$ and other interstitial defects of extrinsic doping, five interstitial sites are randomly generated by inserting the atom into the interstitial space of HD. The configuration with the lowest formation energy is then selected as the representative defect. For example, $C_{i4}$ is the fourth interstitial defect among the five random $C_i$, and it is selected due to its lowest formation energy.

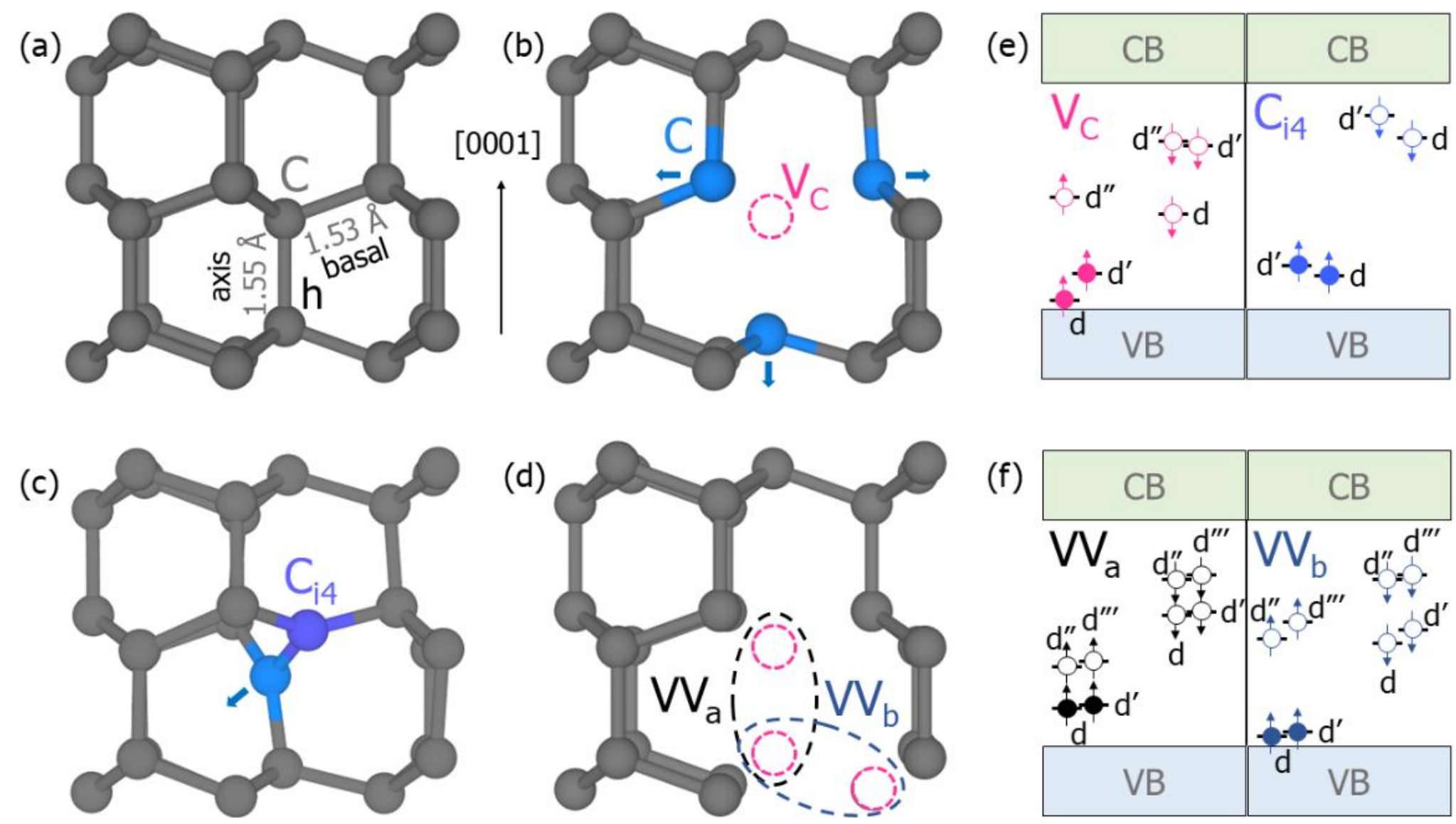

FIG. 1. Structural configurations of (a) host HD, (b) neutral $V_C$, (c) neutral $C_{i4}$, and (d) neutral divacancies $VV_a$, $VV_b$ after relaxation. (e) and (f) represent the calculated defect electronic states of the related defects shown in (b)-(d) with spin polarization.

As shown in Fig. 1(b), the introduction of $V_C$ in HD causes the four neighboring C atoms to move away from the vacancy site, leaving four dangling bonds. These dangling bonds generate four defect electronic states (or defect states). Three of these states, denoted as d, d' and d'', lie within the band gap of HD as shown in Fig. 1(e). The remaining one state resides in the valence band (VB). For the fundamental study of point defects, only the defect states located within the band gap are considered. Thus, we only focus on these states in the following discussion. When spin polarization is included, the three defect states split into six spin states: two occupied spin-up states, one unoccupied spin-up state, and three unoccupied spin-down states. Consequently, $V_C$ can act as a donor by donating up to two occupied electrons, or as an acceptor by receiving up to four electrons into the empty spin states. The calculated formation energy results confirm this analysis. As shown in Fig. 2(a), $V_C$ can be stable in +1, neutral, -1 and -2 charge states depending on the position of the Fermi level ($E_F$), indicating that $V_C$ is a bipolar defect. When acting as a donor, its (1+/0) transition energy level (TEL) is very deep and close to the VBM as shown in Fig. 6(a) or Fig. 2(a). When acting as an acceptor, its (0/1-) TEL lies deep in the middle of the HD band gap, while its (1-/2-) TEL is extremely deep and very close to CBM. These deep TELs result from that the occupied defect states are near the VBM, whereas the unoccupied defect states are near the middle of HD band gap or CBM. Moreover, it is worth noting that the multiple spin states of $V_C$ make it a promising color center for hosting qubits.

The introduction of $C_i$ in HD produces defect states distinct from those of $V_C$ as shown in Fig. 1(e), owing to the different defect configurations. As shown in Fig. 1(c), $C_{i4}$ can lead to significant structural distortion: the interstitial C atom pushes a host C atom (light blue) away from its original lattice site, breaking two original C-C bonds in HD and forming three new C-C bonds between the interstitial C and its neighboring C atoms. This process leaves two dangling bonds, generating two defect states, d and d', within the HD band gap as shown in Fig. 1(e). Considering the spin polarization, these two states split into two occupied spin-up states and two unoccupied spin-down states. Due to the large structural distortion, $C_{i4}$ is highly unstable, as reflected in its relatively high formation energy as shown in Fig. 2(a). For example, the formation energy of neutral $C_{i4}$ is about 12 eV (nearly 5 eV higher than that of neutral $V_C$), and even higher than that of divacancy (VV) defect complexes (discussed below). This indicates that the concentration of $C_{i4}$ is relatively low and its influence on the conductivity of HD is negligible. Moreover, the deep TELs of $C_{i4}$ as shown in Fig. 6(a) also result from the similar reasons to those of $V_C$.

Besides the point defects, the defect complexes are also very important and have been intensively studied in CD[7, 11-13, 27], where they often act as the color centers and host qubits for quantum applications[6], such as the NV center. Therefore, to achieve a comprehensive understanding of defect properties in HD, the defect complexes, specifically the XV complexes (X = V for intrinsic doping and V represents $V_C$; X = Mg, Ca, Sr, B, Al, Ga, Si, Ge, Sn, Pb, N, P and As for extrinsic doping), are also investigated in this work. In intrinsic HD, VV complex consists of two adjacent $V_C$ as shown in Fig. 1(d). Depending on orientation, VV can be classified as $VV_a$ (aligned along the axis direction) and $VV_b$ (aligned along the basal direction). The introduction of VV leaves six dangling bonds, generating six defect states. Among them, four states lie within the HD band gap as shown in Fig. 1(f). As we can see, spin polarization splits these four states into eight spin states: two occupied spin-up states, two unoccupied spin-up states, and four unoccupied spin-down states. These multiple spin states suggest that both $VV_a$ and $VV_b$ hold potential for acting as color centers and hosting the qubits for quantum applications. The distribution of these spin states within the band gap is similar for both $VV_a$ and $VV_b$, and their calculated formation energies are also comparable as shown in Fig. 2(a), reflecting their similar configurations and the comparable distances between adjacent $V_C$. However, due to their relatively high formation energies compared with $V_C$, the influence of these two VV complexes on the conductivity of HD is limited. Hence, in intrinsic HD, the conductivity depends primarily on the $V_C$, and $E_F$ is pinned by the compensation between +1 donor state and -1 acceptor state of $V_C$. As a result, $E_F$ lies approximately 1-1.5 eV above the VBM, leading to the weak p-type conductivity in intrinsic HD.

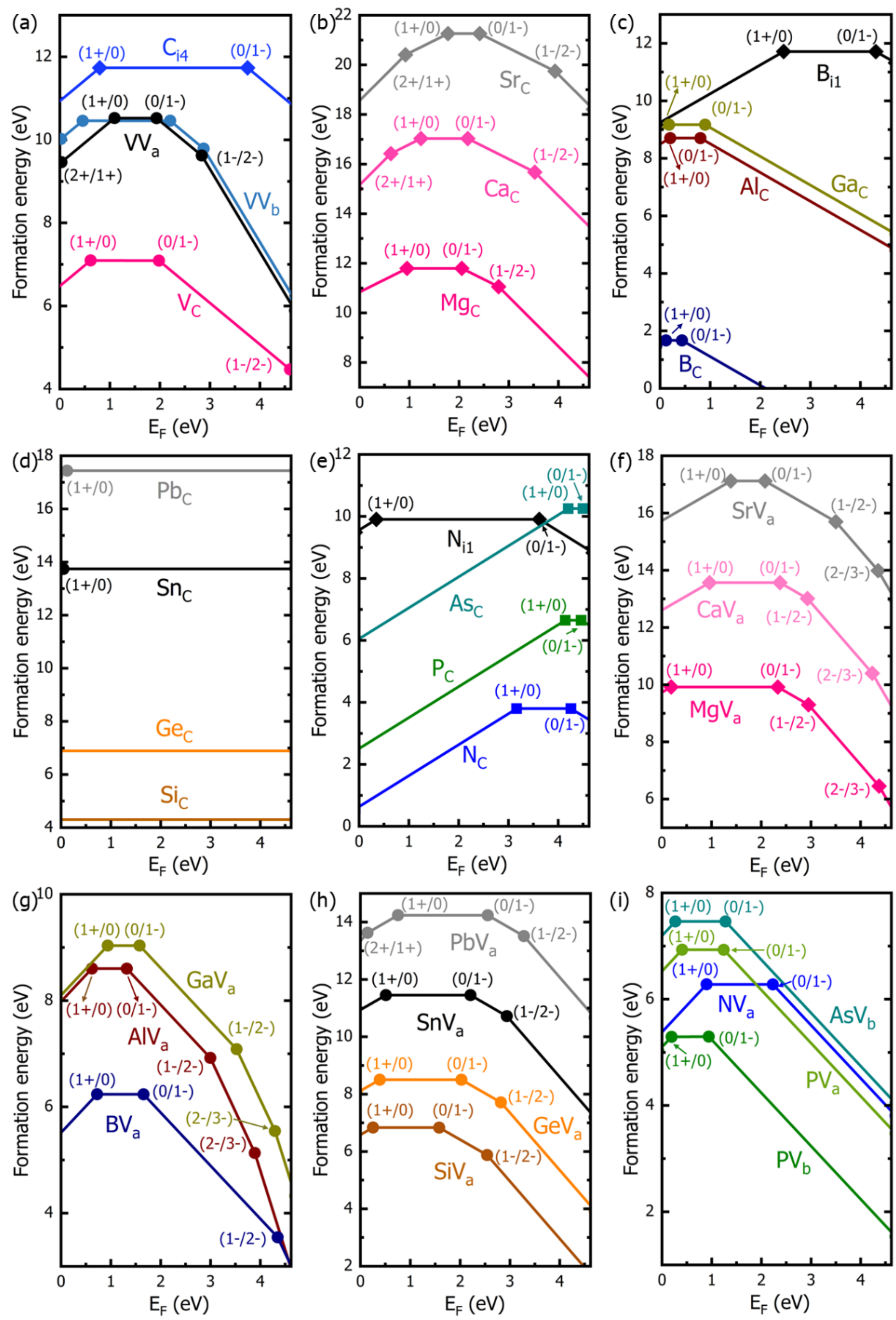

FIG. 2. The calculated formation energies of (a) $V_C$, $C_{i4}$, $VV_a$, $VV_b$, (b) $Mg_C$, $Ca_C$, $Sr_C$, (c) $B_C$, $B_{i1}$, $Al_C$, $Ga_C$, (d) $Si_C$, $Ge_C$, $Sn_C$, $Pb_C$, (e) $N_C$, $N_{i1}$, $P_C$, $As_C$, (f) $MgV_a$, $CaV_a$, $SrV_a$, (g) $BV_a$, $AlV_a$, $GaV_a$, (h) $SiV_a$, $GeV_a$, $SnV_a$, $PbV_a$, (i) $NV_a$, $PV_a$, $PV_b$ and $AsV_b$ defects in HD as function of Fermi level ($E_F$) under C-rich condition. The related chemical potentials of the specific elements are listed in Table S3 of the Supplemental Material[22]. Square, circle and diamond points represent the transition energy levels.

## B . Extrinsic doping of Mg, Ca and Sr

Mg, Ca and Sr are alkaline earth metals belonging to group II of the periodic table. For their extrinsic doping, only substitution defects are considered, while the interstitial defects are neglected. This is because the relatively large atomic sizes of these elements cause significant structural distortion when introduced as interstitials, rendering such configurations unstable. Even for C, which has a smaller atomic size, the intrinsic $C_{i4}$ still has a very high formation energy, making the influence of $C_{i4}$ negligible. Similar to C, the extrinsic B and N doping also have the same situation, where their interstitial defects are also unstable and can be disregarded (as discussed in the following sections). All these make us neglect the interstitial defects of Mg, Ca and Sr doping safely.

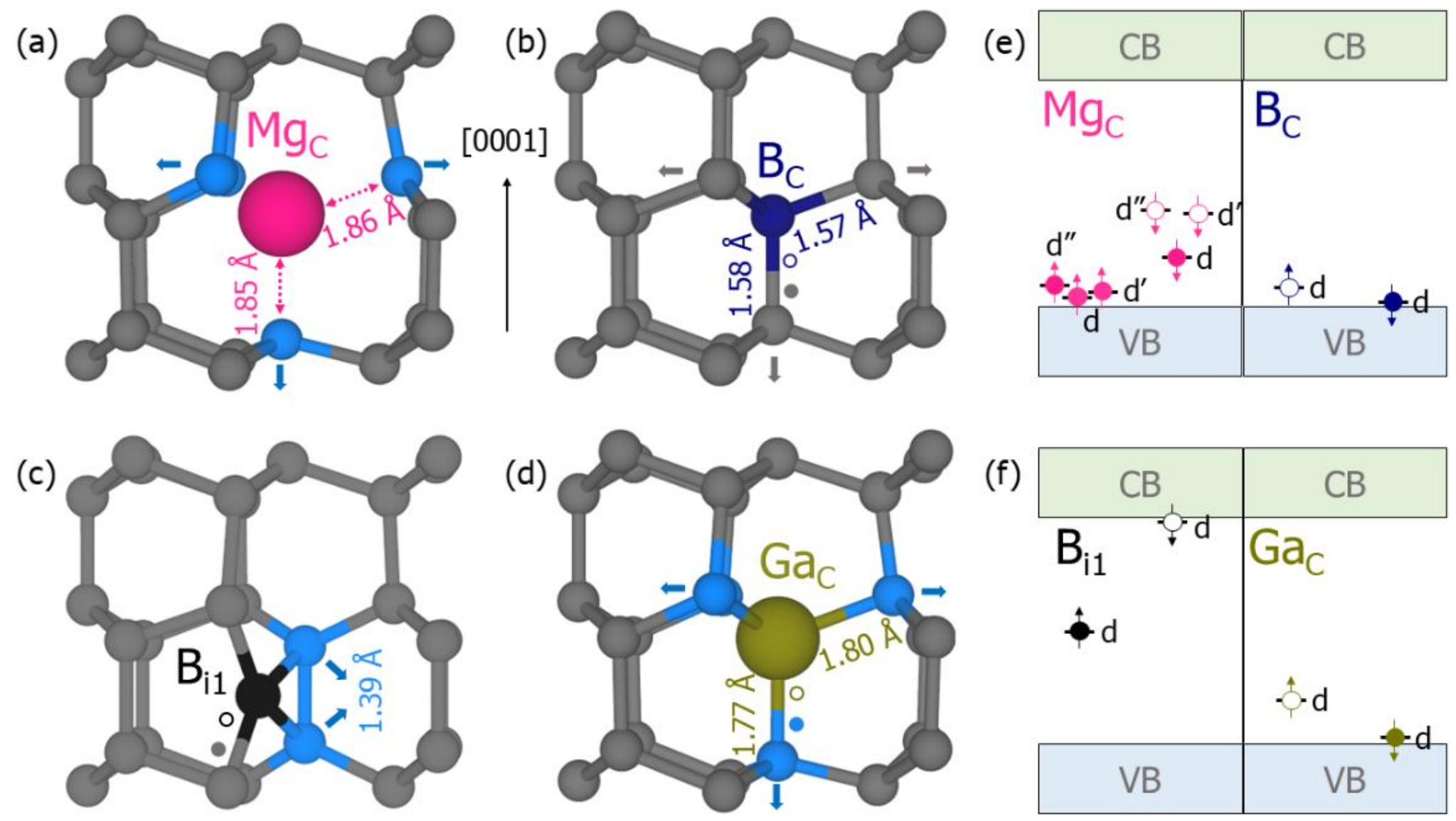

FIG. 3. Structural configurations of (a) neutral $Mg_C$, (b) neutral $B_C$, (c) neutral $B_{i1}$, and (d) neutral $Ga_C$ after relaxation. (e) and (f) represent the calculated defect electronic states of the related defects shown in (a)-(d) with spin polarization.

For their substitution defects, the defect configurations, states, and formation energy trends are broadly similar. For example, $Mg_C$ will significantly push the surrounding C atoms outward as shown in Fig. 3(a), elongating the distance between Mg and C along axis (from 1.55 to 1.85 Å) and basal (from 1.53 to 1.86 Å) directions. Due to the hybridizations of four C dangling bonds and 3s, 3p orbitals of Mg, three defect states (d, d', d'') are introduced within the HD band gap. Spin polarization further splits these states into six spin states, four of which are occupied and the remaining two states are empty as shown in Fig. 3(e). These multiple spin states also make $Mg_C$ a promising color center, similar to the intrinsic $V_C$. The calculated formation energy indicates that $Mg_C$ can be stable in +1, neutral, -1 and -2 charge states depending on the position of $E_F$. Therefore, $Mg_C$ also shows bipolar behavior similar to $V_C$. Specifically, $Mg_C$ acts predominantly as an acceptor across most $E_F$ values, and only when $E_F < 0.95$ eV (the (1+/0) TEL as shown in Fig. 6(a)), it exhibits donor behavior.

For $Ca_C$ and $Sr_C$, they have similar defect configurations to $Mg_C$ as shown in Fig. S1 of the Supplemental Material[22]. However, since Ca and Sr have larger atomic sizes, $Ca_C$ and $Sr_C$ induce more pronounced structural distortions, deviating significantly from ideal $sp^3$ hybridization. These structural distortions result in that the defect states of $Ca_C$ and $Sr_C$ move upward, and an occupied spin-up state flips into the spin-down state for $Sr_C$. Consequently, the (1+/0), (0/1-) and (1-/2-) TELs shift upward as shown in Fig. 6(a), and new (2+/1+) TELs emerge for $Ca_C$ and $Sr_C$. Although $Mg_C$ has the lowest formation energy (near 12 eV for neutral charge state) compared with $Ca_C$ and $Sr_C$, this value is still higher than that of VV complexes (near 10.5 eV for neutral charge states) as shown in Fig. 2. Therefore, these substitution defects of Mg, Ca and Sr doping can not alter the conductivity of HD significantly.

### C . Extrinsic doping of B, Al and Ga

B, Al and Ga belong to the group III elements. In general, they are effective acceptor dopants for HD, especially the B element, which is adjacent to C element and has three valence electrons, one fewer than C. The similar atomic size of B minimizes the structural distortion in the $B_C$ substitution defect as shown in Fig. 3(b). The absence of one valence electron leaves a C dangling bond, introducing a half-occupied defect state in the HD band gap as shown in Fig. 3(e). As shown in Fig. 6(a), the (1+/0) and (0/1-) TELs of $B_C$ lie close to VBM, which is due to the proximity of the empty spin-up and occupied spin-down states to VBM. As shown in Fig. 2(c), $B_C$ exhibits -1 charge state across most $E_F$ values, neutral when $E_F < 0.44$ eV ((0/1-) TEL), and +1 only when $E_F < 0.12$ eV ((1+/0) TEL). Moreover, $B_C$ has a very low formation energy (< 2 eV), confirming its role as a benign acceptor capable of enhancing p-type conductivity in HD.

By contrast, the interstitial B is unstable due to the large structural distortion. As shown in Fig. 3(c), the introduction of $B_{i1}$ breaks two original C-C bonds in host HD and pushes two C atoms (light blue) inward, forming $sp^3$ hybridization and leaving a C dangling bond (a defect state in the HD band gap) as shown in Fig. 3(c) and (f). This structural distortion results in the high formation energy as shown in Fig. 2(c) (near 12 eV for neutral charge state), indicating the instability of $B_{i1}$. Similarly, interstitial defects of Al and Ga are negligible. For substitution defects, $Al_C$ and $Ga_C$ exhibit similar configurations, defect states and formation energy trends to $B_C$, while their larger atomic sizes cause greater distortions, shifting empty spin-up states upward and raising the (0/1–) TELs as shown in Fig. 3 and Fig. 6(a). With the relatively high formation energies (around 9 eV for the neutral charge states), $Ga_C$ and $Al_C$ will not significantly influence the conductivity of HD.

### D . Extrinsic doping of Si, Ge, Sn and Pb

Si, Ge, Sn and Pb are group IV elements, belonging to the same group as C but with larger atomic sizes. As a result, their interstitial defects are more unstable than $C_i$, and only substitution defects are important. However, substitution defects of these elements generally do not contribute to conductivity modification in HD, as they neither offer residual electrons nor holes. The primary motivation for studying these substitution defects is that their formation energies are fundamentally significant for assessing the stability of related XV complexes (X = Si, Ge, Sn, Pb), which are important candidates for color centers in CD[28]. Furthermore, investigating the extrinsic doping of these elements can also enhance the systematic understanding of point defects in HD.

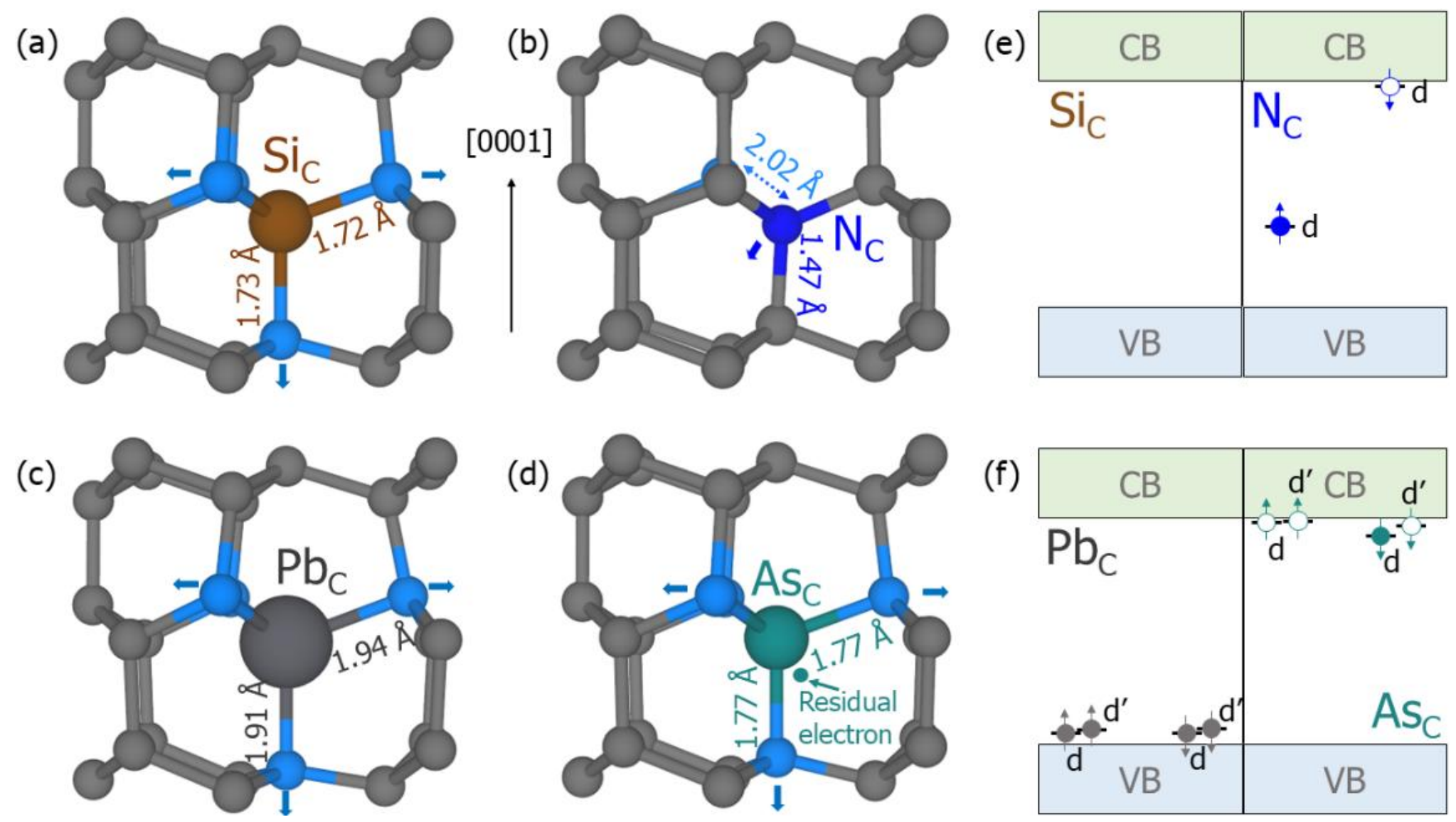

FIG. 4. Structural configurations of (a) neutral Si$_C$, (b) neutral N$_C$, (c) neutral Pb$_C$, and (d) neutral As$_C$ after relaxation. (e) and (f) represent the calculated defect electronic states of the related defects shown in (a)-(d) with spin polarization.

When a Si atom substitutes for a carbon atom (Si$_C$), it typically maintains a neutral charge state as shown in Fig. 2(d) and does not introduce localized defect states within the HD band gap as shown in Fig. 4(e). These are easy to understand. Si has the same outer electronic configuration ($s^2p^2$) as C and can participate in $sp^3$ hybridization, forming four covalent bonds with neighboring C atoms as shown in Fig. 4(a). Although Si is larger than C in atomic size, the Si-C bond lengths (1.73 Å for axis direction and 1.72 Å for basal direction) are sufficiently compatible with the HD lattice without breaking bonds or creating dangling bonds that would lead to defect states in the band gap. Meanwhile, the electronic states of Si are energetically well-aligned with the VBs and conduction bands (CBs) of HD, so any Si-related levels often remain buried within the bulk bands rather than appearing as localized gap states.

Ge$_C$ exhibits similar behavior to Si$_C$, but its larger atomic size leads to greater structural distortion and higher formation energy. Sn$_C$ and Pb$_C$ show even higher formation energies due to their larger atomic sizes. For example, the Pb-C bond lengths (1.91 Å for axis direction and 1.94 Å for basal direction) are significantly longer than those in Si$_C$ as shown in Fig. 4(c) and (a), reflecting pronounced outward displacement of neighboring C atoms and severe lattice distortion. For Sn$_C$ and Pb$_C$, a (1+/0) TEL appears close to the VBM as shown in Fig. 6(a), indicating that it is a very deep donor level. This arises due to the following reasons: Unlike Si or Ge, heavy elements like Pb have much more diffuse orbitals. The energy required to hybridize these orbitals into bonds is higher, and the resulting Pb-C bonds are much weaker (longer bond length). The severe distortion of the lattice as mentioned above and the poor orbital overlap cause some electronic states to be pushed out of the bulk bands and into the band gap. For Pb$_C$, these states manifest as occupied levels just above the VBM as shown in Fig. 4(f), enabling the donor behavior. However, because these occupied states are deep (far from the CB), both Sn$_C$ and Pb$_C$ are deep donors rather than effective shallow donors.

### E . Extrinsic doping of N, P and As

N, P and As are group V elements with one more valence electron than C, making them potential n-type dopants. Although N is smaller than C in atomic size, the interstitial N ($N_i$) in HD is still unstable. For example, $N_{i1}$ has a configuration similar to $C_{i4}$ as shown in Fig. S2 of the Supplemental Material[22]. The introduction of $N_{i1}$ will generate a C dangling bond and a half-occupied defect state in the HD band gap. The occupied spin-up state lies near the VBM, while the empty spin-down state is very close to the CBM, producing deep donor TEL (1+/0) and acceptor TEL (0/1-) as shown in Fig. 6(a). With a high formation energy (near 10 eV for neutral charge state) as shown in Fig. 2(e), $N_{i1}$ has negligible impact on HD conductivity.

For the substitution defect $N_C$, three N-C bonds form and their lengths are all around 1.47 Å, while an original C-C bond is broken as shown in Fig. 4(b). The displaced C atom moves outward (2.02 Å from N), leaving a dangling bond and a half-occupied defect state in the band gap as shown in Fig. 4(e). $N_C$ has a relatively low formation energy (near 4 eV for neutral charge state), which is even lower than that of intrinsic $V_C$ as shown in Fig. 2(e) and (a). $N_C$ is stable in +1 charge state across most $E_F$ values, indicating its dominant donor behavior. Hence, N is a benign n-type dopant, which can increase the concentration of electron carriers effectively.

P and As are larger than N. Therefore, their interstitial defects are neglected due to greater structural distortions. For the substitution defects, $P_C$ and $As_C$ have similar defect configurations and states, yet distinct from those of $N_C$. For example, as shown in Fig. 4(d), $As_C$ forms the standard $sp^3$ hybridization with four neighboring C atoms, using four valence electrons for co-valent bonding (The As-C bond lengths are both 1.77 Å along the axis and basal directions) and leaving one weakly bound residual electron. This residual electron manifests as a localized half-occupied defect state (d state) just below the CBM as shown in Fig. 4(f). Another empty defect state d' arises from the interaction between As and the neighboring C atoms. This interaction creates bonding (filled, in the VB) and anti-bonding (empty, near or in the CB) orbitals. The d' state is one of the anti-bonding states. Formation energy results show that $As_C$ has a high for-mation energy (above 10 eV for the neutral charge state), making its impact negligible, while $P_C$ has a comparable formation energy with that of the intrinsic $V_C$ (both around 7 eV for the neutral charge states), making its impact important. Moreover, since the (1+/0) TEL of $P_C$ is close to the CBM as shown in Fig. 6(a), this means that this TEL is a shallow donor level and $P_C$ can act as an effective donor, enhancing the n-type conductivity in HD, similar to its role in CD[29].

### F . Defect complexes of extrinsic doping

In CD, XV defect complexes (X = B, Si, Ge, Sn, Pb, N) have been intensively studied due to their potential quantum applications[7, 12, 13]. For example, Qiu et al. recently find that the -3 charged MgV center can act as an ideal telecom-band emitter at 1448 nm, and their study provides a long-sought candidate for fiber-compatible quantum emitters[11]. All these investi-gations highlight the importance of XV defect complexes, motivating a systematic study of their properties in HD. As a result, we consider extrinsic dopants X = Mg, Ca, Sr, B, Al, Ga, Si, Ge, Sn, Pb, N, P and As, as mentioned in Sec. A. Similar to intrinsic VV complexes, two orien-tations of XV complexes are considered: $XV_a$ along the axis direction and $XV_b$ along the basal direction. Structural relaxations reveal that XV complexes in HD are stable in two distinct configurations, denoted $XV_I$ and $XV_{II}$ as shown in Fig. 5(a). When X = B and N, the XV complexes adopt the $XV_I$ configuration, in which the extrinsic atom substitutes for the original C atom in HD, forming the $X_C$ substitution defect, and meanwhile an adjacent $V_C$ coexists to complete the defect complex. When X represent the remaining extrinsic elements, the XV complexes adopt $XV_{II}$ configuration. In this configuration, the extrinsic element occupies the interstitial site between two adjacent $V_C$, displacing surrounding C atoms, stabilizing in an equidistant position relative to six neighboring carbons (L1–L6), and forming the split-vacancy configuration V-X-V as shown in Fig. 5(a). This configuration is labeled as XV by the quantum optics groups as mentioned in the previous work[13].

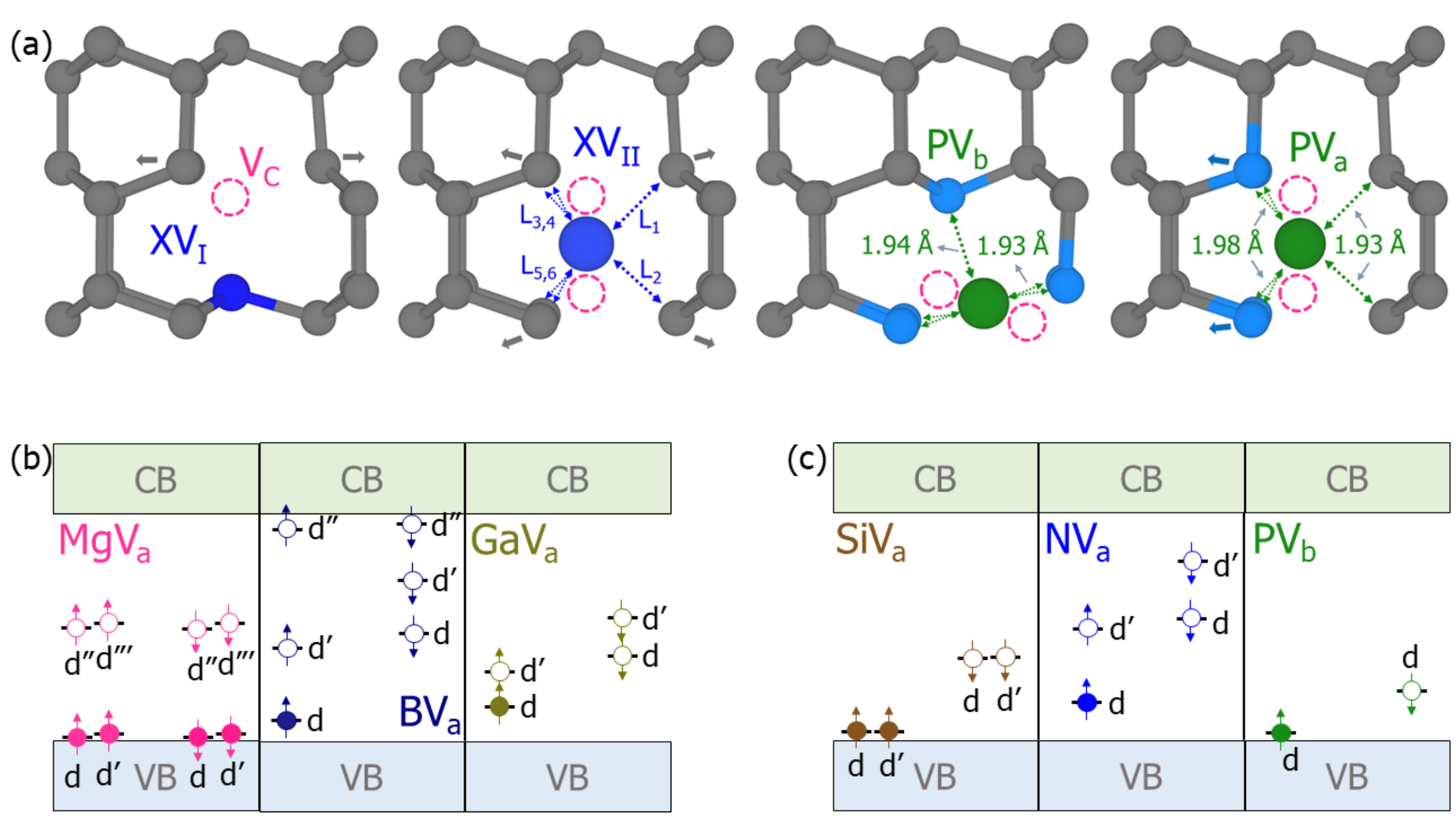

FIG. 5. (a) Structural configurations of $XV_I$ defect complex (X= B, N), $XV_{II}$ defect complex (X = Mg, Ca, Sr, Al, Ga, Si, Ge, Sn, Pb, P, As), and neutral $PV_b$, $PV_a$ after relaxation. (b) and (c) represent the calculated defect electronic states of the selected typical defect complexes ($MgV_a$, $BV_a$, $GaV_a$, $SiV_a$, $NV_a$ and $PV_b$) with spin polarization.

For most extrinsic XV defect complexes, except PV and AsV, the calculated formation energies are comparable for $XV_a$ and $XV_b$, similar to the cases of $VV_a$ and $VV_b$ as shown in Fig. 2(a). As a result, in order to simplify the expression, only the formation energies of $XV_a$ are shown in Fig. 2(f)-(i). For PV, however, the formation energy difference between $PV_a$ and $PV_b$ is about 2 eV as shown in Fig. 2(i), indicating $PV_b$ is more stable. This enhanced stability likely arises from reduced structural distortion in $PV_b$. As shown in Fig. 5(a), the $L_1$ and $L_2$ in $PV_a$ are apparently different from $L_3$-$L_6$, whereas the distances between P and the surrounding C atoms are nearly identical in $PV_b$. AsV exhibits a similar situation, and thus only the formation energy of $AsV_b$ is shown in Fig. 2(i).

Across all calculated formation energies, XV defect complexes display bipolar behavior and tend to be stable in negative charge states for most $E_F$ positions, indicating predominant acceptor characteristics. Notably, the formation energies of $BV_a$, $SiV_a$, $PV_b$, $NV_a$, $PV_a$ and $AsV_b$ are comparable to that of intrinsic $V_C$, suggesting that these complexes may serve as effective p-type dopants and influence hole carrier concentrations in HD. Moreover, as we can see from Fig. 5(b), (c) and Fig. 6(b), except PV and AsV, all the other XV complexes can introduce multiple spin and charge states within the HD band gap. Therefore, these complexes hold huge potential as color centers for quantum applications and deserve further investigations.

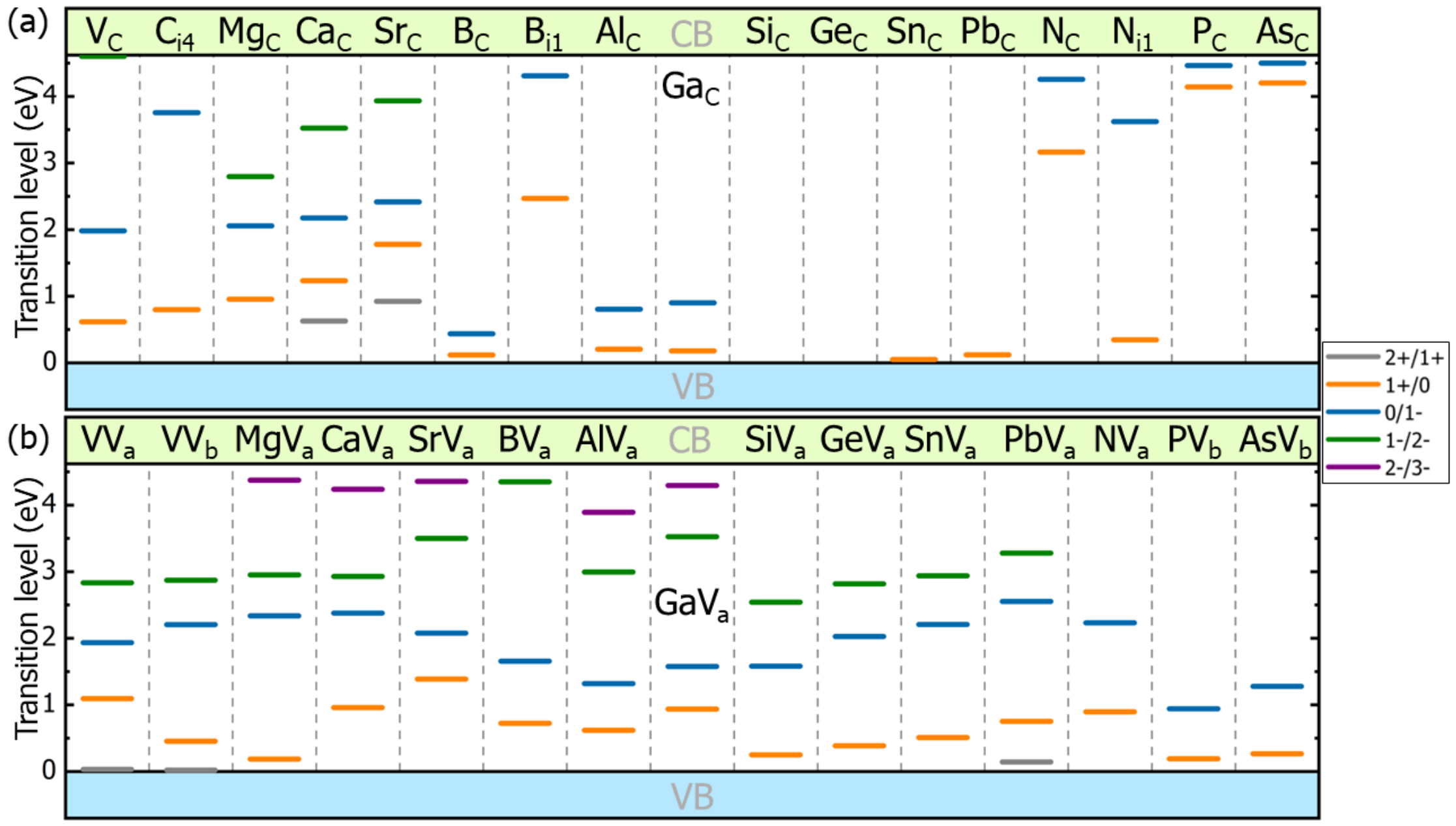

FIG. 6. The calculated transition energy levels of (a) the intrinsic and extrinsic point defects; (b) intrinsic and extrinsic defect complexes in the band gap of HD.

## IV. CONCLUSION

In this work, we have conducted a systematic first-principles investigation of intrinsic and extrinsic point defects, as well as defect complexes, in bulk hexagonal diamond (HD). Our results reveal that intrinsic defects such as carbon vacancies ($V_C$) dominate the conductivity of HD, with the Fermi level pinned by compensation between donor and acceptor states, leading to weak p-type behavior. Carbon interstitials ($C_{i4}$) are highly unstable and thus negligible in conductivity contributions. For extrinsic doping, Group III elements (particularly B) act as effective acceptors, enhancing p-type conductivity, while Group V elements (notably P and N) serve as efficient donors, improving n-type conductivity. In contrast, Group II and Group IV dopants generally show high formation energies or neutral charge states, limiting their impact on carrier concentration. Moreover, $V_C$, $Mg_C$ and XV defect complexes exhibit multiple spin and charge states within the HD band gap, underscoring their potential as color centers for hosting qubits. These findings establish HD as a promising platform for both electronic conductivity engineering and quantum information technologies, warranting further experimental exploration.

## ACKNOWLEDGMENTS

L. Zhu, X. Zhang and G. Alimu contribute equally to this work. This work was supported by National Natural Science Foundation of China (NSFC) under grant Nos. 12304110 and

12404093.